\documentclass[preprint,showpacs,preprintnumbers,amsmath,amssymb]{revtex4}

\usepackage{graphicx}
\usepackage{dcolumn}
\usepackage{bm}

\usepackage{epsfig}

 \def\equal{\begin{array}{c} \mbox{\tiny def} \\[-8pt] = \\[-9pt] {} \end{array}}

 \def\VERT#1{^{|#1}}
 \def\dell{\!\vartriangle\!\!}
 \def\Ai{\mathrm{Ai}}
 \def\Bi{\mathrm{Bi}}

 \def\im{i}
 \def\dd{\mathrm{d}}
 \def\ds{\mathrm{d}s}
 \def\dsigma{\mathrm{d}\sigma}

 \def\K{\mathcal{K}}
 \def\E{\mathcal{E}}
 \def\P{\mathcal{P}}

 \def\P{\mathcal{P}}
 \def\F{\mathcal{F}}
 \def\Fa{\mathcal{F}_\alpha}
 \def\Fb{\mathcal{F}_\beta}

 \def\r{\boldsymbol{r}}
 \def\n{\boldsymbol{n}}
 \def\f{\boldsymbol{f}}
 \def\g{\boldsymbol{g}}
 \def\i{\boldsymbol{i}}
 \def\j{\boldsymbol{j}}
 \def\k{\boldsymbol{k}}
 \def\aa{\boldsymbol{a}}
 \def\bb{\boldsymbol{b}}
 \def\t{\boldsymbol{t}}

 \def\av#1{\langle{#1}\rangle}
 \def\~#1{\widetilde{#1}}
 \def\A{\mathcal{A}}
 \def\abs#1{\left|#1\right|}
 \def\ddA#1{\ddot{\mathcal{A}}_{#1}}

 \def\vs{\vskip 0.5cm}
 
 \def\bfr{{\pmb{r}}}

 \def\^{\wedge}

 \def\bfd{\pmb{d}(\sigma,\eta)}
 
\begin{document}

\title{Gravitational Wave Induced Vibrations\\   of Slender
Structures in Space}

\author{Robin W Tucker}
\altaffiliation[URL: ]{http://www.lancs.ac.uk/depts/imgg/imgg.htm}
 \email{r.tucker@lancaster.ac.uk}
\author{Charles Wang}
\altaffiliation[URL: ]{http://www.lancs.ac.uk/depts/imgg/imgg.htm}
 \email{c.wang@lancaster.ac.uk}
\affiliation{ Department of Physics\\Lancaster University, LA1
4YB, UK}

\begin{abstract} This paper explores the interaction of weak
gravitational fields with slender elastic materials in space  and
estimates their  sensitivities for the detection of gravitational
waves with frequencies between $10^{-4}$ and
 $1$  Hz.
 The dynamic behaviour of such slender structures
is ideally suited to analysis by the simple theory of Cosserat
rods.
 Such a description offers a clean conceptual
separation of the vibrations induced by bending, shear, twist and
extension and the response to gravitational tidal accelerations
can be reliably estimated in terms of the constitutive properties
of the structure. The sensitivity estimates are based on a
truncation of the theory in the presence of thermally induced
homogeneous Gaussian stochastic forces.
\end{abstract}

\maketitle

\section{Introduction}

Gravitational waves are thought to be produced by astrophysical
phenomena ranging from the coalescence of orbiting binaries to
violent events in the early Universe. Their detection would herald
a new window for the observation of natural phenomena. Great
ingenuity is being exercised in attempts to detect such waves in
the vicinity of the earth using either laser interferometry or
various resonant mass devices following Weber's pioneering efforts
with aluminum cylinders. Due to the masking effects of competing
influences and the weakness of gravitation compared with the
electromagnetic interactions the threshold for the detection of
expected gravitationally induced signals remains tantalisingly
close to the limits set by currently technology. In order to
achieve the signal to noise ratios needed for the unambiguous
detection of gravitational waves numerous alternative strategies
are also under consideration. These include more sophisticated
transducer interfaces, advanced filtering techniques and the use
of dedicated arrays of antennae. Earth based gravitational wave
detectors require expensive vibration insulation in order to
discriminate the required signals from the background. This is one
reason  why the use of  antennae in space offer certain
advantages. It is argued here that the gravitationally induced
elastodynamic vibrations of  slender material structures in space
offer other advantages that do not appear to have been considered.
Multiple structures of such continua possess attractive properties
when used as coincidence detectors of gravitational disturbances
with a dominant spectral content in the $10^{-4}$ to  $1$ Hz
region. Furthermore this window can be readily extended to lower
frequencies and higher sensitivities by enlarging the size of the
structures.

Newtonian elastodynamics \cite{antman} is adequate as a first
approximation if supplemented by the {\it tidal stresses}
generated by the presence of spacetime curvature that is small in
comparison with the size of the detector. The latter are estimated
from the accelerations responsible for spacetime geodesic
deviations. Since the constituents of material media owe their
elasticity to primarily non-gravitational forces their histories
are non-geodesic. The geodesic motions of particles offer a
reference configuration and the geodesic deviation of neighbours
in a geodesic reference frame provide accelerations that are
additionally resisted in a material held together by elastic
forces.  Since in practical situations the re-radiation of
gravitational waves is totally negligible the computation of the
stresses induced by the tidal tensor of a background incident
gravitational wave offers a viable means of exploring the
dynamical response of a material domain to a fluctuating
gravitational field.

 Current resonant mode detectors are designed to permit
reconstruction of the direction and polarisation of gravitational
waves that can excite resonances \cite{pp}.  Clearly such
detectors are designed to respond to a narrow spectral window of
gravitational radiation and are not particularly good at
determining the temporal profile of incident gravitational pulses.
A significant advantage of space-based antennae based on slender
material structures is that they can be designed  to respond to
transient gravitational pulses, to polarised uni-directional
gravitational waves or omni-directional unpolarised waves.


The general mathematical theory of non-linear Newtonian elasticity
is well established. The general theory of one-dimensional
Newtonian Cosserat continua derived as limits of three-dimensional
continua can be consulted in \cite{antman}.
 The theory is fundamentally formulated in the
Lagrangian picture in which material elements are labelled by $s$.
 The  behaviour of a Cosserat rod at time $t$ may be described in terms
  of the motion ${\boldsymbol R}(s,t)$ in space of the line of centroids of its cross-sections and
 elastic deformations about that line.  Such a rod  is  modelled
mathematically by an elastic space-curve with structure. This
structure defines the relative orientation of neighbouring
cross-sections along the rod. Specifying a unit vector
${\boldsymbol d}_3$ (which may be identified with the normal to
the cross-section) at each point along the rod enables the state
of shear to be related to the angle between this vector and the
tangent to the space-curve.  Specifying a second vector
${\boldsymbol d}_1$ orthogonal to the first vector (thereby
placing it in the plane of the cross-section) can be used to
encode the state of flexure and twist along its length. Thus a
field of two mutually orthogonal unit vectors along the structure
provides three continuous dynamical degrees of freedom that,
together with the continuous three degrees of freedom describing a
space-curve relative to some arbitrary origin in space, define a
simple Cosserat  model (see Figure 1). It is significant for this
approach that the theory includes thermal variables that can be
coupled to the dynamical equations of motion, compatible with the
laws of thermodynamics. The theory is completed with equations
that relate the deformation strains of the structure to the
elastodynamic forces and torques. The simplest constitutive model
to consider is based on  Kirchoff relations with shear deformation
and viscoelasticity. Such a Cosserat model provides a well defined
six dimensional quasi-linear hyperbolic system of
(integro-)partial differential equations in two independent
variables. It may be applied to the study of gravitational wave
interactions by suitably choosing external body forces
${\boldsymbol f}$ to represent the tidal interaction with each
element in the medium.

A typical system might consist of at least two loops orbiting in
interplanetary space. Each structure would be composed of high Q
material, several km in length.
Such a structure,  
made up of transportable segments,
could be conveyed to an orbiting station and  constructed in
space. The lowest quadrupole excitation of a steel circular
structure would be about 1~Hz and vary inversely as its
(stress-free) length. Actuator and feedback instrumentation could
be placed around the antennae to ``tune'' the system to an optimal
reference configuration. The precise details of the density and
elastic moduli needed to enhance the sensitivity of the receiver
would result from an in-depth analytic analysis of the Cosserat
equations for free motion. The ability to readily optimise the
resonant behaviour for coupled axial, lateral and torsional
vibrations by design is a major advantage over other mechanical
antennae that have been proposed.

 By contrast a broad band
detector could consist of an open ended structure coiled into a
spiral. For planar spirals with traction free open ends the
spectral density of normal transverse and axial linearised modes
increases with the density of the spiral winding number. They form
ideal broad band detectors with directional characteristics.
Furthermore by coupling such a spiral at its outer end to a light
mass by a short length of high-Q fibre (such as sapphire) one may
tune such an extension to internal resonances, thereby amplifying
the spiral elastic excitation. Such excitations offer new
detection mechanisms based on the enhanced motion of the outer
structure of the spiral.

\def\ds{ rod }

\def\bfR{{\boldsymbol r}}
\def\bfn{{\boldsymbol n}}
\def\bfk{{\boldsymbol k}}
\def\bfi{{\boldsymbol i}}
\def\bfj{{\boldsymbol j}}

\def\bff{{\boldsymbol f}}
\def\bfw{{\boldsymbol w}}
\def\bfm{{\boldsymbol m}}
\def\bfl{{\boldsymbol l}}
\def\bfdk{\,{\boldsymbol d}^{(k)}}
\def\bfdi{\,{\boldsymbol d}^{(i)}}
\def\bfd{\,{\boldsymbol d}}
\def\bfr{{\boldsymbol r}}
\def\bft{{\boldsymbol t}}
\def\bfN{{\boldsymbol N}}
\def\FNB{{{\cal F}{}^{bit}_N}(\eta)}
\def\FF{{\cal F}}

\def\bfF{{\boldsymbol F}}
\def\bfJ{{\boldsymbol J}}
\def\bfL{{\boldsymbol L}}
\def\bfQ{{\boldsymbol Q}}
\def\bfV{{\boldsymbol V}}

\def\bfLQ{{\boldsymbol L_Q}}

\def\bfv{{\boldsymbol v}}
\def\bfu{{\boldsymbol u}}

 \vskip 0.5cm

\centerline{\bf Cosserat Equations} \vskip 0.5cm

\def\hh{\hfill\break}
\def \d {{\bfd}}
\def \vs {\vskip 0.4cm}
\def \r {{\bfR}}
\def \n {{\bfn}}
\def \m {{\bfm}}
\def \u {{\bfu}}
\def \v {{\bfv}}
\def \f {{\boldsymbol f}}
\def \II  {{\boldsymbol I}}
\def \R {{\bfR}}
\def\cross{\times}
\def\bfomega{\bfw}

\def \st {(s,t)}
\def \sst {(s_0,t)}

  The dynamical evolution of a\ds with mass density, $s\in
[0,L]\mapsto\rho(s),$ and cross-sectional area, $s\in [0,L]\mapsto
A(s),$ is governed by Newton's dynamical laws: $$ \rho A\,
\ddot\r\,=\n^\prime+\f\label{1} $$ $$
\partial_t(\rho\,\II(\bfomega))= \m^\prime+\r^\prime\cross\n +
\bfl\label{2} $$ applied to a triad of orthonormal vectors: $s\in
[0,L]\mapsto\{ {\boldsymbol d_1} (s,t),{\boldsymbol d_2}
(s,t),{\boldsymbol d_3} (s,t)\}$ over the space-curve: $s\in
[0,L]\mapsto\r (s,t)$ at time $t$ where
$\n^\prime=\partial_s\n,\,\, \dot\r=\partial_t\r$, $\bff$ and
$\bfl$ denote external force and torque densities respectively and
$s\in [0,L]\mapsto\rho\II$ is a \ds moment of inertia tensor. In
these field equations the {\it contact forces} $\n$ and {\it
contact torques} $\m$ are related  to the vector $\bfomega$ and
the  {\it strains} $\u,\,\v$ by constitutive relations. These
vectors are themselves defined in terms of the configuration
variables $\r$ and ${\d_k\,\,}$ for $ k=1,2,3$ by the relations:
$$ \r^\prime=\v\label{5}$$
$$ \d_k^\prime=\u\cross\d_k$$
 $$\dot\d_k={\bfomega}\cross\d_k.$$
  The latter ensures that the triad
remains orthonormal under evolution.  The last equation identifies
$$ {\bfw}= \frac{1}{2} \sum_{k=1}^3 {\bfd_k} \cross
{\dot{\bfd_k}}$$ with the local angular velocity vector of the
director triad. The general model accommodates continua whose
characteristics (density, cross-sectional area, rotary inertia)
vary with $s$. For a system of two coupled continua with different
elastic characteristics on $0 \le s < s_0$ and $s_0 < s \le L$
respectively one matches the degrees of freedom at $s=s_0$
according to a junction condition describing the coupling.

To close the above equations of motion constitutive relations
appropriate to the \ds must be specified: $ \bfn(s,t)=
\hat\bfn(\bfu(s,t),\bfv(s,t),\bfu_t(s),\bfv_t(s),\ldots,s) $, $
\bfm(s,t)=
\hat\bfm(\bfu(s,t),\bfv(s,t),\bfu_t(s),\bfv_t(s),\ldots,s) $ where
$\bfu_t(s)$ etc., denote the history of $\bfu(s,t)$ up to time
$t$. These relations specify a  reference configuration (say at
$t=0$) with strains ${\boldsymbol U}(s),{\boldsymbol V}(s)$ such
that $ \hat\bfn({\boldsymbol U}(s),{\boldsymbol V}(s),\ldots,s) $
and $ \hat\bfm({\boldsymbol U}(s),{\boldsymbol V}(s),\ldots,s) $
are specified. A  reference configuration  free of flexure has
$\bfR_s(s,0)=\bfd_3(s,0)$, i.e. $ {\boldsymbol V(s)}=\bfd_3(s,0).
$ If a standard configuration is such that $\bfR(s,0)$ is a
space-curve with Frenet curvature $\kappa_0$ and torsion $\tau_0$
and the standard directors are oriented so that $\d_1(s,0)$ is the
unit normal to the space-curve and $\bfd_2(s,0)$ the associated
unit binormal then $ {\boldsymbol U}(s)=
\kappa_0(s)\,\bfd_2(s,0)+\tau_0(s)\,\bfd_3(s,0). $


\def\F{{\cal F}}
For a rod of density $\rho$ and cross-sectional area $A$ in a weak
plane gravitational wave background the simplest model to consider
consists of the Newtonian Cosserat equations above with a time
dependent body force modelled by a gravitational  tidal
interaction ${\boldsymbol f}$. In addition to time-dependent waves
this may include  stationary Newtonian gravitational fields. These
add terms of the form $ \rho A \tilde{\boldsymbol g}$ to
${\boldsymbol f}$ where $\tilde{\boldsymbol g}$ is the ``effective
local acceleration due to gravity''. Post Newtonian gravitational
fields (such as gravitomagnetic and Lens-Thirring effects) can be
accommodated with a more refined metric background.

\def\LLL{\frac{L}{2\pi}}
\def\LL{\frac{2\pi}{L}}
\def\a{\alpha}
\def\b{\beta}

An important consideration of any modelling of Cosserat continua
to low levels of excitation is the estimation of  signal to noise
ratios induced by  anelasticity and thermal interactions. To gain
an insight into the former one may attempt to extend the
established theory of linear anelasticity to a Cosserat structure.
For a string with uniform density $\rho$, static Young's modulus
$E$ and area of cross section $A$, the  free damped motion in one
dimension is modelled   by the equation:
$$\rho A \,\partial_{tt} x(s,t) = n^\prime(s,t)$$ where the axial
strain $v(x,t)=\partial_s x(s,t)\equiv x^\prime(s,t)$ and

$$ n(s,t)=EA(v(s,t)-1)-EA\int_{-\infty}^t \phi(t-t^\prime)\,\dot
v(s,t^\prime) d t^\prime$$ for some viscoelastic model $\phi$ with
$0\le s \le L$.
 For free motion in the mode:
$$ x(s,t)=s+\xi(t) \cos(\pi s/L)$$
the amplitude $\xi(t)$ satisfies
$$\ddot\xi(t)+\omega_0^2\, \xi(t)=\omega_0^2 \int_{-\infty}^t
\phi(t-t^\prime)\dot \xi(t^\prime) d t^\prime$$ with
$\omega_0^2=\frac{\pi^2 E}{L^2\rho}$ while for a forced harmonic
excitation:
$$\ddot\xi(t)+\omega_0^2 \xi(t)=\omega_0^2\, \int_{-\infty}^t
\phi(t-t^\prime)\dot \xi(t^\prime) d t^\prime+F_0 \exp(-i \Omega
t). $$  With $\xi(0)=\dot\xi(0)=0$ the Laplace transformed
amplitude of forced axial motion is then given in terms of the
Laplace transform \footnote{For a "Hudson " type solid \cite{hud}
: $\bar E(\sigma)=k \sigma^\nu =E(1-\sigma \bar\phi(\sigma))$ for
some constants $k$ and $\nu$.}

$$\bar\phi(\sigma)=\int_0^{\infty}\phi(t) e^{-\sigma t}
d\,t $$ of the anelastic modelling function  $\phi(t)$ as:
$$\bar\xi(\sigma)=\frac{F_0}{(\sigma- i\Omega)(\sigma^2
-\omega_0^2 \sigma\bar\phi(\sigma)+ \omega_0^2) }.$$

\vskip 0.5cm

 To extend this approach in a simple manner to a  3-D Cosserat
model  of a slender rod with uniform static moduli $E$ and $G$,
geometric elements $A$, $K_{\a\a}=J_{11}+ J_{22}$, $J_{\a\b}$, one
adopts the following constitutive relations for the local director
components of the contact force ${\boldsymbol n}$ and torque
${\boldsymbol m}$ in terms of the local strain vectors
${\boldsymbol v}$ and ${\boldsymbol u}$ and anelastic response
functions $\phi_E$ and $\phi_G$:

$$n_3(s,t)=EA\,(v_3(s,t)-1)- EA\,\int_{-\infty}^t
\phi_E(t-t^\prime)\, \dot v_3(s,t^\prime)\,d\, t^\prime$$

$$n_1(s,t)=GA\,v_1(s,t)- GA\,\int_{-\infty}^t \phi_G(t-t^\prime)\,
\dot v_1(s,t^\prime)\,d\, t^\prime$$

$$n_2(s,t)=GA\,v_2(s,t)- GA\,\int_{-\infty}^t \phi_G(t-t^\prime)\,
\dot v_2(s,t^\prime)\,d\, t^\prime$$

$$m_3(s,t)=K_{\a\a}G\,u_3(s,t)- K_{\a\a}G\,\int_{-\infty}^t
\phi_G(t-t^\prime)\, \dot u_3(s,t^\prime)\,d\, t^\prime$$

$$m_\a(s,t)=E\sum_{\b=1}^{2}J_{\a\b}\,u_\b(s,t)-
E\sum_{\b=1}^{2}J_{\a\b}\,\int_{-\infty}^t \phi_E(t-t^\prime)\,
\dot u_\b(s,t^\prime)\,d\, t^\prime$$ for $\a,\b=1,2$.

\section{Behaviour of an elastic string in a noisy  weak gravitational wave background}

The above outlines a new approach to the modelling of
gravitational interactions with slender structures in space. Given
the topology and material properties of an antenna one may analyse
its response to such signals in terms of solutions to a
deterministic  system of well defined partial
(integro-)differential equations. These equations are in general
easier to analyse than those describing the elastodynamics of
three-dimensional materials. However in addition to  controlling
the temperature dependence of material characteristics, thermal
interactions with such structures will induce a stochastic element
into the signal response. It is therefore necessary to seek
modifications to the above Cosserat description that can
accommodate such random interactions.  In the presence of internal
damping this is non-trivial problem  for a broad band resonant
detector.

In the calculation that follows we shall oversimplify this problem
in order to gain some order of magnitude estimates of
signal-to-(thermal)noise ratios for both narrow and broad band
resonant detectors made of known high-Q materials. The simplest
approach is to approximate the Cosserat equations by ignoring
flexure and torsional mode excitations and explore the purely
string-like excitations. Commensurate with these approximations we
shall assume that the damping can be described in terms of a
single Q-factor at about $4$ K for each resonant mode and that the
thermal interactions give rise to a stochastic system driven by  a
spatially homogeneous Gaussian noise term.

We shall make two further restrictions by assuming that the
antenna possesses sufficient stiffness to maintain planar motion
in tension (e.g. by overall slow rotation about its centre of
mass) throughout the excitation time and that the environment
produces no overall translational drift away from a regular
orbital motion of its centroid. The plane will be chosen
orthogonal to the direction of propagation of a gravitational
disturbance.

We therefore begin by examining the linearised modes about an
arbitrary planar 1-dimensional structure. In this way we can
address the response of both a narrow band planar loop and a broad
band planar spiral together. Given the anelastic characteristics
of the structure all of these restrictions can be readily relaxed
at the cost of increased complexity in the thermo-mechanical
analysis.

In such a framework consider the (inertial) Cartesian basis
$\{\i,\j,\k\}$ with corresponding coordinates $(x,y,z)$ such that
the tidal acceleration field at any position
$\r=x\,\i+y\,\j+z\,\k$ and time $t$, corresponding to a plane
gravitational wave travelling in the direction  $\k$ at the speed
of light $c$ has the form \cite{schutz}
\begin{eqnarray}
\g(\r,t) &=& \left\{\ddA{1}(t-z/c)\,x +
\ddA{2}(t-z/c)\,y\right\}\frac{\i}{2} + \left\{\ddA{2}(t-z/c)\,x -
\ddA{1}(t-z/c)\,y\right\}\frac{\j}{2}.
\label{vecf}
\end{eqnarray}
 The dimensionless waveforms
$\A_1(t-z/c)$ and $\A_2(t-z/c)$ arise from two independent
polarisations of the gravitational wave in the
transverse-traceless gauge in linearised gravitation. This
acceleration excites an elastic string with reference length $L$,
cross-section area $A$ and mass density $\rho$ described by a
space-curve $\r(s,t)$ parameterised by the  parameter $s \in [0,
L]$ according to
\begin{eqnarray}
{\rho A}\,\ddot{\r}(s,t) - \n'(s,t) &=& {\rho A}\,\g(\r(s,t),t).
\label{motion}
\end{eqnarray}
The  contact force (tension) $\n(s,t)$ is given by the
constitutive relation:
\begin{eqnarray}
\n(s,t) &=& E A \left(\left|\r'(s,t)\right| - 1\right)\,
\frac{\r'(s,t)}{\left|\r'(s,t)\right|}
\end{eqnarray}
in terms of the Young's modulus $E$. For a closed string, the
periodic boundary conditions are
\begin{eqnarray}
\r(0,t) &=& \r(L,t)
\label{bcnd_closed1}
\\[5pt]
\r'(0,t) &=& \r'(L,t).
\label{bcnd_closed2}
\end{eqnarray}
 For an open string we  consider the tension-free boundary
conditions given by
\begin{eqnarray}
\n(0,t) &=& \n(L,t) = 0.
\label{bcnd_open}
\end{eqnarray}
Let $\r_0(s)$ be a stress-free {\it reference configuration} of
the elastic string with a unit stretch, i.e. $\abs{\r_0'(s)}=1$.
Assuming the Frenet curvature of $\r_0(s)$, given by
\begin{eqnarray}
\kappa_0(s) &\equal& |\r_0''(s)|
\label{curv}
\end{eqnarray}
is non-vanishing  introduce a Frenet frame as the set of
orthonormal vectors along the string: \cite{struik}
\begin{eqnarray}
\t_0(s) &\equal& \r_0'(s)
\label{t}
\\[5pt]
\aa_0(s) &\equal& \frac{\t_0'(s)}{\kappa_0(s)}
\label{a}
\\[5pt]
\bb_0(s) &\equal& \t_0(s) \times \aa_0(s).
\label{b}
\end{eqnarray}
The  space-curve of the reference configuration $\r_0(s)$ is
defined in the $x$-$y$ plane with $z=0$ so that
\begin{eqnarray}
\r_0(s) &=& x_0(s)\,\i + y_0(s)\,\j.
\label{r0}
\end{eqnarray}
With
\begin{eqnarray}
\bb_0 &=& \k
\label{bk}
\end{eqnarray}
it follows  \cite{struik} that the tangent vector $\t_0(s)$ may be
expressed in terms of the Frenet curvature of the reference
space-curve as
\begin{eqnarray}
\t_0(s)
&=& -\sin \left( \int_0^s \kappa_0({s'}) \, \dd {s'} + \theta
\right) \i + \cos \left( \int_0^s \kappa_0({s'}) \, \dd {s'} +
\theta \right) \j
\label{tij}
\end{eqnarray}
for an arbitrary constant $\theta$. Let $\alpha(s,t)$ and
$\beta(s,t)$ be axial and transverse perturbations in the $x$-$y$
plane of the string about $\r_0(s)$ so that
\begin{eqnarray}
\r(s,t) &=& \r_0(s) + \alpha(s,t)\,\t_0(s) + \beta(s,t)\,\aa_0(s).
\label{r_perturb}
\end{eqnarray}
Since the interaction between the string and the gravitational
wave is assumed small we shall drop terms containing powers of
$\alpha,\beta,\A_1,\A_2$ higher than 1 in the following. Thus
\eqref{vecf} becomes
\begin{eqnarray}
\g(\r(s,t),t) &=& \left\{ \ddA{1}(t)\,F_1(s) + \ddA{2}(t)\,F_2(s)
\right\}\t_0(s) \nonumber
\\[5pt]
&& + \left\{ \ddA{2}(t)\,F_1(s) - \ddA{1}(t)\,F_2(s)
\right\}\aa_0(s)
\label{vecff1}
\end{eqnarray}
up to first order terms in $\alpha$ and $\beta$, where
\begin{eqnarray}
F_1(s) &\equal& \frac{1}{4} \left\{ x_0(s)^2-y_0(s)^2 \right\}'
\label{F1}
\\[5pt]
F_2(s) &\equal& \frac{1}{2} \left\{ x_0(s)\,y_0(s) \right\}'.
\label{F2}
\end{eqnarray}
>From (\ref{t}), (\ref{a}), (\ref{b}) and (\ref{bk})
\begin{eqnarray}
\r' &=& \left( 1 + \alpha' - \kappa_0\,\beta \right)\t_0 + \left(
\alpha\,\kappa_0 + \beta' \right)\aa_0
\label{expd_r}
\end{eqnarray}
and therefore
\begin{eqnarray}
\n' &=& E A \left( \alpha' - \kappa_0\,\beta \right)' \t_0 + E A
\, \kappa_0 \left( \alpha' - \kappa_0\,\beta \right) \aa_0
\label{pn}
\end{eqnarray}
up to first order terms in $\alpha$ and $\beta$. Substituting this
into (\ref{motion}) to first order yields
\begin{eqnarray}
\ddot\alpha(s,t) -\frac{E}{\rho}\,\left\{\alpha'(s,t) -
\kappa_0(s)\,\beta(s,t)\right\}' &=& \ddot{\Fa}(s,t)
\label{eqn_alpha}
\\[5pt]
\ddot\beta(s,t) -\frac{E}{\rho}\,\kappa_0(s)\,\left\{\alpha'(s,t)
- \kappa_0(s)\,\beta(s,t)\right\} &=& \ddot{\Fb}(s,t)
\label{eqn_beta}
\end{eqnarray}
where
\begin{eqnarray}
\Fa(s,t) &\equal& \A_{1}(t)\,F_1(s) + \A_{2}(t)\,F_2(s)
\label{Fa}
\\[5pt]
\Fb(s,t) &\equal& \A_{2}(t)\,F_1(s) - \A_{1}(t)\,F_2(s).
\label{Fb}
\end{eqnarray}
For an open string the tension-free condition \eqref{bcnd_open}
becomes
\begin{eqnarray}
\alpha'(0,t) - \kappa_0(0)\,\beta(0,t) = \alpha'(L,t) -
\kappa_0(L)\,\beta(L,t) &=& 0.
\label{bcnd1}
\end{eqnarray}
Introducing
\begin{eqnarray}
\chi(s,t) &\equal&
\alpha(s,t)-\left(\frac{\beta(s,t)}{\kappa_0(s)}\right)'
\label{def_chi}
\end{eqnarray}
and using (\ref{eqn_alpha}) and (\ref{eqn_beta}) gives
\begin{eqnarray}
\ddot \chi(s,t) &=& \ddot\Fa(s,t) -
\left(\frac{\ddot\Fb(s,t)}{\kappa_0(s)} \right)'
\label{eqn_chi}
\end{eqnarray}
which implies
\begin{eqnarray}
\chi(s,t) &=& \Fa(s,t) - \left(\frac{\Fb(s,t)}{\kappa_0(s)}
\right)' +\chi_0(s)+\chi_1(s)\,t
\label{sol_chi}
\end{eqnarray}
for two arbitrary functions $\chi_0(s)$ and $\chi_1(s)$.
Furthermore by introducing
\begin{eqnarray}
\xi(s,t) &\equal&
\frac{\beta(s,t) -\Fb(s,t)}{L\,\kappa_0(s)}
\label{def_xi}
\end{eqnarray}
and substituting (\ref{def_chi}) and \eqref{sol_chi} into
(\ref{eqn_beta}) we obtain
\begin{eqnarray}
\ddot\xi(s,t) -\frac{E}{\rho}\,\left\{ \xi''(s,t) -
\kappa_0(s)^2\,\xi(s,t)\right\} &=& \frac{1}{L}\, \frac{E}{\rho}
\left\{ \F(s,t) +\chi_0'(s)+\chi_1'(s)\,t \right\}
\label{eqn_xi}
\end{eqnarray}
where
\begin{eqnarray}
\F(s,t) \equal \Fa'(s,t) -\kappa_0(s)\,\Fb(s,t) &=&
\A_{1}(t)\,\F_1(s) + \A_{2}(t)\,\F_2(s)
\label{F}
\end{eqnarray}
in terms of
\begin{eqnarray}
\F_1(s) &\equal& F_1'(s)+\kappa_0(s)\,F_2(s) = -\frac{1}{2}\,\cos
\left( 2\K_0(s) + \theta \right)
\label{FFF1}
\\[5pt]
\F_2(s) &\equal& F_2'(s)-\kappa_0(s)\,F_1(s) = -\frac{1}{2}\,\sin
\left( 2\K_0({s}) + \theta \right)
\label{FFF2}
\end{eqnarray}
with
\begin{eqnarray}
\K_0(s) &\equal& \int_0^s \kappa_0({s'}) \, \dd {s'}.
\label{K0}
\end{eqnarray}
In deriving the above relations, (\ref{tij}), (\ref{F1}) and
(\ref{F2}) have been used. It follows from \eqref{def_chi},
\eqref{sol_chi} and \eqref{def_xi} that
\begin{eqnarray}
\alpha(s,t) &=& L\,\xi'(s,t) + \Fa(s,t) +\chi_0(s)+\chi_1(s)\,t
\label{alpha_sol}
\\[5pt]
\beta(s,t) &=& L\,\kappa_0(s)\xi(s,t) + \Fb(s,t).
\label{beta_sol}
\end{eqnarray}
For an open string with tension-free ends we may substitute
\eqref{alpha_sol} and \eqref{beta_sol} into (\ref{bcnd1}),
(\ref{def_chi}) to obtain
\begin{eqnarray}
\ddot\xi(0,t) = \ddot\xi(L,t) &=& 0
\label{bcnd_xi_open}
\end{eqnarray}
by using  \eqref{eqn_xi}. Therefore $\alpha(s,t)$ and $\beta(s,t)$
can be obtained by solving \eqref{eqn_xi} for a choice of
functions $\chi_0(s)$ and $\chi_1(s)$, subject to the boundary
condition \eqref{bcnd_xi_open} for an open string or
\begin{eqnarray}
\xi(0,t) &=& \xi(L,t)
\label{bcnd_xi_closed1}
\\ [5pt]
\xi'(0,t) &=& \xi'(L,t)
\label{bcnd_xi_closed2}
\end{eqnarray}
together with $\chi_0(0) = \chi_0(L), \chi_0'(0) = \chi_0'(L)$ and
$\chi_1(0) = \chi_1(L), \chi_1'(0) = \chi_1'(L)$ for a closed
string.

\section{Normal mode analysis}

We next analyse \eqref{eqn_alpha} and \eqref{eqn_beta} for
displacement perturbations that remain small compared with $L$ at
all times. These solutions represent the deterministic dynamic
response of a string in terms of small oscillatory deviations from
a time-independent reference configuration $\r_0(s)$ under the
influence of weak gravitational waves. Such $\alpha(s,t)$ and
$\beta(s,t)$ can be obtained by choosing
\begin{eqnarray}
\chi_0(s)=\chi_1(s) &=& 0
\label{no_drift}
\end{eqnarray}
that enter \eqref{eqn_xi}, \eqref{alpha_sol} and \eqref{beta_sol}.
For an open string we further choose
\begin{eqnarray}
\xi(0,t)=\xi(L,t) &=& 0
\label{no_drift1}
\end{eqnarray}
as solutions to \eqref{bcnd_xi_open}. To proceed we express
$\xi(s,t)$ in terms of normal modes,
 a complete set of basis
functions $\{\psi_p(s)\}$ indexed by (integer- or multi-integer
valued) $p$ satisfying
\begin{eqnarray}
\psi_p''(s) + \left\{ \frac{\rho}{E}\,\omega_p^2 - \kappa_0(s)^2
\right\} \psi_p(s) &=& 0
\label{eigen_phi}
\end{eqnarray}
subject to either periodic or tension-free
$(\psi_p(0)=0,\,\psi_p(L)=0)$ boundary conditions with the
corresponding eigen-values $\{\frac{\rho}{E}\,\omega_p^2\}$. In
addition the eigen-functions $\psi_p(s)$ shall satisfy the
ortho-normality conditions
\begin{eqnarray}
\int_0^L {\psi_p(s)\,\psi_q(s)}\,{\rm d} s &=& L\,\delta_{pq}.
\label{orth_phi}
\end{eqnarray}
In this  basis
\begin{eqnarray}
\xi(s,t) &=& \sum_{p} \xi_p(t)\,\psi_p(s)
\label{exp_p}
\end{eqnarray}
summing over the full range of $p$.

Substituting \eqref{no_drift} and (\ref{exp_p}) into
(\ref{eqn_xi}) and using (\ref{eigen_phi}) and (\ref{orth_phi})
gives
\begin{eqnarray}
\ddot{\xi}_p(t) + \omega_p^2\,\xi_p(t) &=& f_p(t)
\label{eqn_xip}
\end{eqnarray}
where
\begin{eqnarray}
f_p(t) &\equal& \frac{1}{L}\, \frac{E}{\rho} \left\{ a_{1
p}\,\A_{1}(t) + a_{2 p}\,\A_{2}(t) \right\}
\label{fp}
\end{eqnarray}
with the ``overlap coefficients'' given by
\begin{eqnarray}
a_{1 p} &\equal& \frac{1}{L}\,\int_0^L \F_1(s)\,\psi_p(s)\,\dd s
\label{c1}
\\[5pt]
a_{2 p} &\equal& \frac{1}{L}\,\int_0^L \F_2(s)\,\psi_p(s)\,\dd s.
\label{c2}
\end{eqnarray}
It follows from (\ref{FFF1}) and (\ref{FFF2}) that these
coefficients do not depend on the choice of the origin on the
$x$-$y$ plane. In addition the coefficients
\begin{eqnarray}
a_p &\equal& \sqrt{a_{1 p}^2+a_{2 p}^2}
\end{eqnarray}
are independent of $\theta$ and  invariant under rotation of the
$x$- and $y$-axes about the $z$-axis. The value of $a_p$ provides
a measure of the ``coupling strength'' between the $p$-th-normal
mode and the exciting  gravitational wave.

\section{Mechanical energy in terms of normal modes}

The mechanical energy of the string  satisfying
\eqref{motion} is given by 
\begin{eqnarray}
\hat\E &=& \frac{1}{2}\, \int_0^L \left\{ \rho A\, \dot{\r}^2 + E
A \left(\abs{\r'} - 1\right)^2 \right\} \dd s.
\label{mechen}
\end{eqnarray}
Substituting (\ref{r_perturb}) into (\ref{mechen}) gives
\begin{eqnarray}
\hat\E &=& \frac{1}{2}\, \int_0^L \left\{ \rho A\,
\left(\dot{\alpha}^2+\dot{\beta}^2\right) + E A \left(\alpha' -
\kappa_0\,\beta\right)^2 \right\} \dd s \label{mechen1}
\end{eqnarray}
up to first order terms in $\alpha$ and $\beta$. By using
\eqref{alpha_sol} and \eqref{beta_sol} the above expression
becomes
\begin{eqnarray}
\hat\E &=& \frac{1}{2}\, \int_0^L \left\{ {\rho A} \left(
L^2\,\dot{\xi}'{}^2 + L^2\,\kappa_0\,\dot{\xi}^2 +
\dot{\F}_\alpha^2 + \dot{\F}_\beta^2 - 2\,L\,\dot{\xi}\,\dot{\F}
\right) + {E A} \left( L\,\xi'' - L\,\kappa_0^2\,\xi + \F
\right)^2 \right\} \dd s. \nonumber
\end{eqnarray}
In the absence of the gravitational wave excitation
this reduces to
\begin{eqnarray}
\E &=& \frac{{\rho A\,L^2}}{2}\, \int_0^L \left\{
\dot{\xi}'{}^2+\kappa_0\,\dot{\xi}^2 + \frac{\rho}{E} \,
\ddot{\xi}^2 \right\} \dd s
\label{mechen3}
\end{eqnarray}
where (\ref{eqn_xi}) has been used. From (\ref{exp_p}),
(\ref{eigen_phi}) and (\ref{orth_phi}) together with boundary
conditions \eqref{bcnd_xi_open} for an open string or
\eqref{bcnd_xi_closed1}, \eqref{bcnd_xi_closed2} for a closed
string we obtain
\begin{eqnarray}
\E &=& \sum_p \E_p
\label{mechen5}
\end{eqnarray}
with modal contributions
\begin{eqnarray}
\E_p &\equal& \frac{1}{2}\,m_p \left( \omega_p^2\,{\xi}_p^2 +
\dot{\xi}_p^2 \right)
\label{Ep}
\end{eqnarray}
where
\begin{eqnarray}
m_p &\equal& \frac{\rho^2 A\,L^3\,\omega_p^2}{E}
\label{mp}
\end{eqnarray}
may be identified as an effective modal mass.

\section{Sensitivity Estimation}


 To
accommodate the thermal fluctuations and dissipation for a
vibrating string with small displacements, the modal equations
\eqref{eqn_xip} should be modified into a set of {\em coupled
linear stochastic equations}  of the form
\begin{eqnarray}
\ddot{\xi}_p(t) + \sum_{q} 2\,\zeta_{p
q}\,{\omega_q}\,\dot{\xi}_q(t) + \omega_p^2\,\xi_p(t) &=& f_p(t) +
\dot{w}_p(t)
\label{stoch_eq1}
\end{eqnarray}
where the $\{\zeta_{p q}\}$ are (in general non-diagonal) coupling
constants related to material visco-elasticity and  the
$\{{w}_p(t)\}$ are independent Wiener processes characterised by
Gaussian probability densities:
\begin{eqnarray}
\P(w_p(t)-w_p(t-\dell t)) &=& \frac{1} {\sqrt{2 \pi {b}_p \dell
t}} \,\exp{\left\{ -\frac{ \left[ w_p(t)-w_p(t-\dell t) \right]^2}
{2 \,{b}_p  \dell t} \right\}}
\end{eqnarray}
at any time $t$ for some width parameters ${b}_p$. As discussed
above we assume here that $\zeta_{p q}$ for $p \neq q$, vanish
identically ({\em proportional damping} \cite{thomson}). Then
\eqref{stoch_eq1} reduces to  a system of {\em decoupled linear
stochastic equations}
\begin{eqnarray}
\ddot{\xi}_p(t) + 2\,\zeta_p\,{\omega_p}\,\dot{\xi}_p(t) +
\omega_p^2\,\xi_p(t) &=& f_p(t) + \dot{w}_p(t)
\label{stoch_eq}
\end{eqnarray}
in terms of the modal damping ratios $\zeta_p \equal \zeta_{p p}$
for any $p$. These are related to quality factors $Q_p$ and
relaxation times $\tau_p$ by
\begin{eqnarray}
\zeta_p &=& \frac{1}{2 Q_p} = \frac{1}{2 \omega_p \tau_p}.
\end{eqnarray}
To maintain thermal equilibrium at  a temperature $T$ between
fluctuations and dissipation depending on $\zeta_p$ and $w_p(t)$
the parameters ${b}_p$ are given by \cite{chandrasekhar}
\begin{eqnarray}
{b}_p &=& { \frac{2\, k_B T}{m_p \tau_p} }
\end{eqnarray}
involving Boltzmann's constant $k_B$ ($=1.38\times10^{-23}$ J/K.)

Given initial conditions for $\xi_p(t)$ and $\dot{\xi}_p(t)$ at
$t=0$, a representative solution of (\ref{stoch_eq}) for $t > 0$
may be conveniently expressed as
\begin{eqnarray}
\xi_p(t) &=& \xi_{0 p}(t)+\xi_{f_p}(t)+\xi_{\dot{w}_p}(t)
\label{xi_decomp}
\end{eqnarray}
where
\begin{eqnarray}
\xi_{0 p}(t) &\equal& \xi_p(0)\, \sqrt{1-\zeta_p^2}\,\omega_p\,
\Theta_p(t) + (\xi_p(0)\,\zeta_p\,\omega_p + \dot\xi_p(0))\,
\Phi_p(t)
\label{xi_0p}
\\[5pt]
\xi_{f_p}(t) &\equal& \int_0^{\infty} \Phi_p(t-{t'}) f_p({t'}) \,
\dd{t'}
\label{xif}
\\[5pt]
\xi_{\dot{w}_p}(t) &\equal& \int_0^{\infty} \Phi_p(t-{t'})
\dot{w}_p({t'}) \, \dd{t'}
\label{xidw}
\end{eqnarray}
in terms of
\begin{eqnarray}
\Theta_p(t) &\equal& H(t)\,
\frac{e^{-\zeta_p\,\omega_p\,t}}{\sqrt{1-\zeta_p^2}\,\omega_p}\,
\cos({\sqrt{1-\zeta_p^2}\,\omega_p\,t})
\\[5pt]
\Phi_p(t) &\equal& H(t)\,
\frac{e^{-\zeta_p\,\omega_p\,t}}{\sqrt{1-\zeta_p^2}\,\omega_p}\,
\sin({\sqrt{1-\zeta_p^2}\,\omega_p\,t})
\end{eqnarray}
using the Heaviside function
\begin{eqnarray}
H(t) &\equal& \left\{
\begin {array}{cl}
1&\mbox{if}\;\; t \ge 0
\\\noalign{\medskip}
0&\mbox{otherwise.}
\end {array}
\right.
\end{eqnarray}
It follows \cite{chandrasekhar}  that the $\xi_{\dot{w}_p}(t)$
given in \eqref{xidw} give rise to the  time-dependent deviation
values:
\begin{eqnarray}
\av{\xi_{\dot{w}_p}(t)^2} =\frac{k_B T}{m_p \,\omega_p^2} \left\{
1- e^{-t/\tau_p} \Gamma_{+} \right\}
\label{thermal_dis}
\end{eqnarray}

\begin{eqnarray}
\av{\dot{\xi}_{\dot{w}_p}(t)^2} &=& \frac{k_B T}{m_p} \left\{ 1-
e^{-t/\tau_p} \Gamma_{-} \right\}
\label{thermal_vel}
\end{eqnarray}
where $$ \Gamma_{\pm}= 1 \pm
\frac{\zeta_p}{\sqrt{1-\zeta_p^2}}\sin({2\sqrt{1-\zeta_p^2}\,\omega_p\,t})
+
\frac{2\zeta_p^2}{1-\zeta_p^2}\sin^2({\sqrt{1-\zeta_p^2}\,\omega_p\,t}).
$$

For given gravitational waveforms $\A_{1}(t)$ and
$\A_{2}(t)$ and the corresponding $f_p(t)$ evaluated using
\eqref{fp}, the ``signal'' displacements $\xi_{f_p}(t)$ and their
derivatives at any measurement time $t = \tau > 0$ may be
calculated using \eqref{xif} and  compared with those due to
thermal ``noise'' given by \eqref{thermal_dis} and
\eqref{thermal_vel}.

The above noise terms have small variances if $\tau \ll \tau_p$,
as pointed out in \cite{hawking} in analysing a bar-type
gravitational wave detector. In this case for ``high-$Q$''
materials with $\zeta_p \ll 1$ the ratio of ``effective mechanical
energy'' due to signals to ``effective thermal energy'' at
$t=\tau\ll \tau_p$ may be approximated as follows. Based on
\eqref{Ep} introduce the effective  mechanical energy associated
with the displacement $\xi_{f_p}(t)$ in \eqref{xif} due to a
signal at measurement time $\tau$ given by:
\begin{eqnarray}
\E_p^{(S)}({\tau}) &\equal& \frac{1}{2}\,m_p\,{\omega_p^2}\,
{\xi_{f_p}}({\tau})^2 + \frac{1}{2}\,m_p\,
\dot{\xi}_{f_p}({\tau})^2
\label{ES1}
\\[5pt]
&=&
m_p\,\int_0^{\tau} \dot{\xi}_{f_p}({t})\,f({t})\, \dd{t} -
2\,\zeta_p\,{\omega_p}\, m_p\,\int_0^{\tau}
\dot{\xi}_{f_p}({t})^2\, \dd{t}.
\label{ES2}
\end{eqnarray}
Let
\begin{eqnarray}
f_p\VERT{{\tau}}({t}) &\equal& H({\tau}-t)\,f_p(t)
\label{FF}
\end{eqnarray}
and denote accordingly
\begin{eqnarray}
\xi_{f_p\VERT{{\tau}}}(t) &\equal& \int_0^{\infty} \Phi_p(t-{t'})
f_p\VERT{{\tau}}({t'}) \, \dd{t'}
\label{xif1}
\end{eqnarray}
so that $\xi_{f_p}(t)=\xi_{f_p\VERT{{\tau}}}(t)$ for $0 < t <
{\tau}$. Now
\begin{eqnarray}
\int_0^{\tau} \dot{\xi}_{f_p}({t})\,f_p({t})\, \dd{t} &=&
\int_0^{\infty}
\dot{\xi}_{f_p\VERT{{\tau}}}({t})\,{f}_p\VERT{{\tau}}({t})\,
\dd{t}
= \frac{1}{2\pi}\, \Re \int_{-\infty}^{\infty} \im\,\omega\,
\~{\xi}_{f_p\VERT{{\tau}}}(\omega)\,\~{{f}}_p
\VERT{{\tau}}(\omega)^*\, \dd\omega
\end{eqnarray}
in terms of the one-sided Fourier transform operator $\;\~{}\;$
and its inverse given by
\begin{eqnarray}
\~F(\omega) &\equal&
\int_0^{\infty} F(t)\,e^{-\im\omega t} \,\dd t, \;\;\; F(t) =
\frac{1}{2\pi}\, \int_{-\infty}^{\infty} F(\omega)\,e^{\im\omega
t} \,\dd \omega
\label{def_ft}
\end{eqnarray}
for any function $F$ of non-negative $t$. It follows from
(\ref{xif1}) that
\begin{eqnarray}
\~{\xi}_{f_p\VERT{{\tau}}}(\omega) &=& \~{\Phi}_p(\omega)
\~{{f}}_p \VERT{{\tau}}(\omega)
\end{eqnarray}
where
\begin{eqnarray}
\~{\Phi}_p(\omega) &=&
\frac{1}{\omega_p^2-\omega^2+2\,\im\,\zeta_p\,\omega_p\,\omega}
\end{eqnarray}
and therefore
\begin{eqnarray}
\int_0^{\tau} \dot{\xi}_{f_p}({t})\,f_p({t})\, \dd{t} &=&
\frac{1}{2\pi}\, \int_{-\infty}^{\infty}
\frac{2\,\zeta_p\,\omega_p\,\omega^2}
{(\omega_p^2-\omega^2)^2+4\,\zeta_p^2\,\omega_p^2\,\omega^2}\,
\abs{\~{{f}}_p \VERT{{\tau}}(\omega)}^2\, \dd\omega.
\label{ES3}
\end{eqnarray}
For sufficiently small $\zeta_p$
\begin{eqnarray}
\frac{2\,\zeta_p\,\omega_p\,\omega^2}
{(\omega_p^2-\omega^2)^2+4\,\zeta_p^2\,\omega_p^2\,\omega^2}
&\approx& \frac{\pi}{2} \left\{
\delta(\omega-\omega_p)+\delta(\omega+\omega_p) \right\}
\label{approx_delta}
\end{eqnarray}
(in the sense of distributions) and the second term in \eqref{ES2}
becomes negligible. This yields
\begin{eqnarray}
\E_p^{(S)}({\tau}) &\approx& \frac{m_p}{2}\, \abs{\~{{f}}_p
\VERT{{\tau}}(\omega_p)}^2 = \frac{1}{2} \, E A L \, \omega_p^2
\abs{ a_{1 p}\,\~{{\A}}_1\VERT{{\tau}}(\omega_p) + a_{2
p}\,\~{{\A}}_2\VERT{{\tau}}(\omega_p) }^2.
\label{Eps}
\end{eqnarray}
The effective thermal noise energy due to $\xi_{\dot{w}_p}(t)$ in
\eqref{xidw} is given by
\begin{eqnarray}
\E_p^{(N)}({\tau}) &\equal& \frac{1}{2}\,m_p\,{\omega_p^2}\,
\av{\xi_{\dot{w}_p}({\tau})^2} + \frac{1}{2}\,m_p\,
\av{\dot{\xi}_{\dot{w}_p}({\tau})^2}
\\[5pt]
&\approx& 2\,\zeta_p\,\omega_p\, {\tau}\,k_B T =
\frac{{\tau}}{\tau_p}\,k_B T
\label{ES}
\end{eqnarray}
for $\tau\ll \tau_p$ and $\zeta_p \ll 1$. In these limits a
gravitational signal to thermal noise ratio can be estimated to be
\begin{eqnarray}
\frac{\E_p^{(S)}({\tau})}{\E_p^{(N)}({\tau})} &\approx&
\frac{\omega_p^2}{2} \, \frac{{\tau_p}}{\tau} \, \frac{E A L}{k_B
T} \, \abs{ a_{1 p}\,\~{\A}_1\VERT{{\tau}}(\omega_p) + a_{2
p}\,\~{\A}_2\VERT{{\tau}}(\omega_p) }^2.
\label{SN1}
\end{eqnarray}

The response of any antennae to a general pulse of gravitational
radiation can be estimated once its spectral content is known.
Consider then  a linearly polarised harmonic gravitational wave
with the waveforms:
\begin{eqnarray}
\A_{1}(t) &=&
h \sin\omega_g t
\\[5pt]
\A_{2}(t) &=& 0,
\end{eqnarray}
frequency $\omega_g/2\pi$ and dimensionless amplitude $h$. If the
measurement time corresponds to $n$ $(\in \mathbb{Z}^+)$ cycles of
this signal i.e. ${\tau}=\tau_n\equal2 n \pi/\omega_g \ll \tau_p$
then
\begin{eqnarray}
\~\A_{1}\VERT{{\tau_n}}(\omega) &=&
\frac{h\,\omega_g}{\omega^2-\omega_g^2} \left(
e^{-i{\omega}{\tau_n}}-1 \right).
\end{eqnarray}
Hence $\~\A_{1}\VERT{{\tau_n}}(\omega_p)$ has a maximum modulus at
resonance when $\omega_g = \omega_p$, yielding
\begin{eqnarray}
\~\A_{1}\VERT{{\tau_n}}(\omega_g) &=& -\frac{i h n \pi}{\omega_g}.
\label{ftawg}
\end{eqnarray}
In this case the signal-to-noise ratio \eqref{SN1} becomes
\begin{eqnarray}
\frac{\E_p^{(S)}({\tau_n})}{\E_p^{(N)}({\tau_n})} &\approx&
\frac{n \pi h^2 a_{1 p}^2 \, Q_p }{4} \, \frac{E A L}{k_B T}
\label{SN2}
\end{eqnarray}
for an integer $n \ll Q_p/2\pi$. This formula is applicable to any
1-dimensional Cosserat structure with a specified reference
configuration satisfying the material characteristics assumed
above.

\section{A narrow-band circular loop gravitational antenna}
\label{circ}

A circular reference configuration of a closed string may be
parameterised with
\begin{eqnarray}
x_0(s) &=& R\cos\frac{s}{R}
\\[5pt]
y_0(s) &=& R\sin\frac{s}{R}
\\[5pt]
z_0(s) &=& 0.
\end{eqnarray}
By using (\ref{curv}) this has a Frenet curvature
\begin{eqnarray}
\kappa_0(s) &=& \frac{1}{R}.
\end{eqnarray}
The ``shapes'' of the associated normal modes satisfying
(\ref{orth_phi}) and (\ref{eigen_phi}) subject to periodic
boundary conditions are given by
\begin{eqnarray}
\psi_{[k,1]}(s) &=& \sqrt{2}\,\cos\frac{k s}{R}
\label{cpsi1}
\\[5pt]
\psi_{[k,2]}(s) &=& \sqrt{2}\,\sin\frac{k s}{R}
\label{cpsi2}
\end{eqnarray}
with multiple mode indices ($p = [k,1]$ or $[k,2]$ with positive
integer $k$). The associated eigen-angular frequencies are
\begin{eqnarray}
\omega_{[k,1]} = \omega_{[k,2]} &=&
\frac{1}{R}\,\sqrt{(k^2+1)\,\frac{E}{\rho}}.
\label{omega_k12}
\end{eqnarray}
The effective modal masses, obtained by substituting the above
into (\ref{mp}),  are
\begin{eqnarray}
m_{[k,1]} = m_{[k,2]} &=& 8\pi^3\rho A R\,(k^2+1).
\label{cmp}
\end{eqnarray}
It follows from (\ref{FFF1}) and (\ref{FFF2}) with the choice
$\theta=0$ that
\begin{eqnarray}
\F_1(s) &=& -\frac{1}{2}\,\cos\frac{2 s}{R}
\label{cFFF1}
\\[5pt]
\F_2(s) &=& -\frac{1}{2}\,\sin\frac{2 s}{R}.
\label{cFFF2}
\end{eqnarray}
Substituting these into (\ref{c1}) and (\ref{c2}) we obtain the
following overlap coefficients:
\begin{eqnarray}
a_{1,[k,j]} &=& -\frac{1}{2\sqrt{2}}\,\delta_{2,k}\,\delta_{1,j}
\label{cc1}
\\[5pt]
a_{2,[k,j]} &=& -\frac{1}{2\sqrt{2}}\,\delta_{2,k}\,\delta_{2,j}.
\label{cc2}
\end{eqnarray}
Therefore only two quadrupole  modes (corresponding to $p = [2,1]$
and  $[2,2]$) are excited by the gravitational waves. The signal
to thermal noise ratios for these two modes follow from
(\ref{SN1}) as
\begin{eqnarray}
\frac{\E_{[2,j]}^{(S)}({\tau})}{\E_{[2,j]}^{(N)}({\tau})}
&\approx& \frac{\pi\, \omega_{[2,j]}^2}{8} \,
\frac{\tau_{[2,j]}}{{\tau}}\, \frac{E A R}{k_B T}
\abs{\~{\A}_j\VERT{{\tau}}(\omega_{[2,j]})}^2
\label{csn}
\end{eqnarray}
where $j = 1,2$. From \eqref{SN2}
\begin{eqnarray}
\frac{\E_{[2,1]}^{(S)}({\tau_n})}{\E_{[2,1]}^{(N)}({\tau_n})}
&\approx& \frac{n \pi^2 h^2 Q_{[2,1]} }{16} \, \frac{E A R}{k_B
T}.
\label{csn1}
\end{eqnarray}
For $R = 1.8 \times 10^{3}$ m, $\rho = 8 \times 10^{3}$ kg/m$^3$,
$E = 2.0 \times 10^{11}$ kg/m s$^2$, the quadrupole mode frequency
is $\omega_{[2,1]}/2\pi = 1$ Hz. For a structure with a circular
cross-section of radius $r=0.01$ m the area $A = \pi r^2$. If one
selects   $T=4$ K, $Q_{[2,1]} = 10^{5}$ and $n \approx 0.1 \times
Q_{[2,1]}/2\pi$ then the condition
${\E_{[2,1]}^{(S)}({\tau_n})}/{\E_{[2,1]}^{(N)}({\tau_n})} = 1$
implies an amplitude sensitivity of $h \approx 2 \times 10^{-21}$
for such an antenna.

\section{A broad-band spiral gravitational antenna}
\label{spiral}

By contrast to the narrow-band loop  antenna in which only two
quadrupole modes with the same frequency are excited by a harmonic
gravitation wave as discussed in Section~\ref{circ}, we now
demonstrate how a multi-frequency gravitational antenna may be
constructed based on a spiral reference configuration. Consider an
open string whose reference configuration lies on the $x$-$y$
plane as before but with a non-constant Frenet curvature of the
form
\begin{eqnarray}
\kappa_0(s) &=& \frac{1}{R}\,\sqrt{\mu^3\frac{s}{R}+1}
\label{s_curv}
\end{eqnarray}
for a constant $\mu$ and scale parameter $R$. It follows from
\eqref{K0} that
\begin{eqnarray}
\K_0(s) &=& \frac{2}{3}\frac{R^3}{\mu^3} \left[
\kappa_0(s)^3-\kappa_0(0)^3
\right].
\label{sp_int}
\end{eqnarray}
Substituting (\ref{s_curv}) into  (\ref{eigen_phi}) yields
\begin{eqnarray}
\psi_p''(s) + \left( \frac{\mu^2}{R^2}\,\lambda_p -
\frac{\mu^3}{R^3}\,s \right) \psi_p(s) &=& 0
\label{s_eigen_phi}
\end{eqnarray}
where
\begin{eqnarray}
\lambda_p &\equal& \frac{1}{\mu^2} \left(
\frac{\rho}{E}\,R^2\,\omega_p^2 - 1 \right).
\label{def_lambda}
\end{eqnarray}
Solving \eqref{s_eigen_phi} subject to tension-free boundary
conditions yields
\begin{eqnarray}
\psi_p(s) &=& {A_p}\,\Ai\left(\mu\frac{s}{R} - \lambda_p\right) +
{B_p}\,\Bi\left(\mu\frac{s}{R} - \lambda_p\right)
\label{sp_psi}
\end{eqnarray}
and the condition
\begin{eqnarray}
\left [
\begin {array}{cc}
\Ai(-\lambda_p)  & \Bi(-\lambda_p)
\\\noalign{\medskip}
\Ai(\mu \ell-\lambda_p) & \Bi(\mu \ell-\lambda_p)
\end {array}
\right ] \left [\begin {array}{c} A_p
\\\noalign{\medskip}
B_p
\end {array}\right ]
&=& \left [\begin {array}{c} 0
\\\noalign{\medskip}
0
\end {array}\right ]
\label{sp_eigen}
\end{eqnarray}
where $\ell \equal L/R$ and the $\Ai,\Bi$ are the standard Airy
functions. The corresponding characteristic equation for
$\lambda_p$,
\begin{eqnarray}
\Ai(-\lambda_p)\,\Bi(\mu \ell-\lambda_p) - \Ai(\mu
\ell-\lambda_p)\,\Bi(-\lambda_p) &=& 0,
\label{sp_secular}
\end{eqnarray}
determines the eigen-values  $\lambda_p$. When comparing the
spectrum of the eigen-frequencies of a spiral with that of a
circular loop it is convenient to introduce the function
\begin{eqnarray}
\Omega(s) &\equal& \kappa_0(s)\,\sqrt{5\,\frac{E}{\rho}}.
\label{def_Omega}
\end{eqnarray}
Note that for any ${s} \in [0, L]$ the spiral has a radius of
curvature $1/\kappa_0({s})$ and from \eqref{omega_k12} it can be
seen that $\Omega({s})$ equals the quadrupole mode angular
frequency of a circular loop of radius $1/\kappa_0({s})$ with the
same mass density $\rho$ and Young's modulus $E$. Given values of
$\lambda_p$ as solutions of \eqref{sp_secular}, it follows from
\eqref{def_lambda} and \eqref{def_Omega} that the eigen-angular
frequencies $\omega_p$ are given by
\begin{eqnarray}
{\omega_p} &=& {{\Omega(0)}}\, \sqrt{ \frac{\mu^2\,\lambda_p +
1}{5}}
\label{omega_p_Omega}
\end{eqnarray}
and from  (\ref{sp_eigen})
\begin{eqnarray}
B_p &=& -\frac{\Ai(\mu \ell-\lambda_p)}{\Bi(\mu
\ell-\lambda_p)}\,A_p.
\label{sp_eq_B}
\end{eqnarray}
The normalisation condition (\ref{orth_phi}) now takes the form
\begin{eqnarray}
\int_0^\ell \left\{ A_p\,\Ai\left(\mu\sigma-\lambda_p\right) +
B_p\,\Bi\left(\mu\sigma-\lambda_p\right) \right\}^2 \dsigma &=&
\ell
\label{sp_norm}
\end{eqnarray}
where $ \sigma \equal {s}/{R} $. Therefore \eqref{sp_eq_B} and
\eqref{sp_norm} determine $A_p$ and $B_p$. Furthermore the overlap
coefficients $a_{1 p}$ and $a_{2 p}$ are obtained from (\ref{c1})
and (\ref{c2}) using (\ref{FFF1}), (\ref{FFF2}) and
(\ref{sp_psi}).

Finally from  \eqref{SN2} the signal to (thermal) noise ratio for
the $p$-th mode of the spiral  may be written:
\begin{eqnarray}
\frac{\E_p^{(S)}({\tau_n})}{\E_p^{(N)}({\tau_n})} &\approx&
\frac{n \pi \ell h^2 a_{1 p}^2 \, Q_p }{4} \, \frac{E A R}{k_B T}.
\label{scn2}
\end{eqnarray}

In Figure 4 we have plotted, for a given spiral geometry
(determined by the parameters $\mu, \ell$ and $R$, the values of
$\hat{a}_p^2 = \ell (a_{1 p}^2 + a_{2 p}^2)$ for the first 60
normal modes of the spiral according to the equations above. The
modes are distributed uniformly on the abscissa in terms of the
non-dimensional eigen-angular frequencies $\hat{\omega}_p =
\omega_p/\Omega(0)$. The broad-band response feature of this
antenna is clearly visible in the region  where significant signal
overlap occurs.

For $\mu = 0.3$, $\ell = 16\,\pi$, $R = 1.9 \times 10^{3}$ m,
$\rho = 8 \times 10^{3}$ kg/m$^3$, $E = 2.0 \times 10^{11}$ kg/m
s$^2$, one finds for $p=45$ that $\lambda_p = 95.5$ and $a_{1 p} =
0.135$ corresponding to the eigen-frequency $\omega_p/2\pi = 1$
Hz. For a structure with a circular cross-section of radius
$r=0.01$ m the area $A = \pi r^2$. If one selects   $T=4$ K, $Q_p
= 10^{5}$ and $n \approx 0.1 \times Q_{p}/2\pi$ then the condition
${\E_p^{(S)}({\tau_n})}/{\E_p^{(N)}({\tau_n})} = 1$ implies an
amplitude sensitivity of $h \approx 2 \times 10^{-21}$ for such an
antenna. Thus the spiral antenna maintains a sensitivity
commensurate with that of the single loop antenna but with the
added broad-band characteristic.

\section{Conclusions}

The simple Cosserat theory of rods has been outlined and used to
estimate the thermal noise sensitivity of gravitational antennae
constructed out of orbiting slender material structures. Although
such calculations  have not included many other  competing noise
sources we feel that they provide support for a novel concept.
Orbiting Cosserat structures can accommodate both narrow-band and
broad-band detectors and may offer a much cheaper alternative to
existing space-based laser-interferometers. One may venture
optimism that costs will become ever more competitive with the
current rapid development of high-strength carbon-based fibres.
Not withstanding economic considerations, the implementation of
these ideas would be complimentary to existing global programmes
of gravitational wave research. The design of detectors with
optimised response in the 1 Hz  spectral region would exploit a
current niche in the existing gravitational antenna frequency
spectrum and the detection of gravitational waves in this domain
would provide vital information about stochastic backgrounds in
the early Universe and the relevance of super-massive black holes
to  processes  in the centre of galaxies.

Although we have concentrated on planar structures and short
measurement times, in order to gain enhanced sensitivities, the
extension to non-planar loops and spirals is in principle
straightforward. For example,  several \lq\lq wire-balls" of
approximately spherical shape may offer a viable method to monitor
stochastic gravitational waves over longer periods of time.

There remain many  further important issues that need detailed
attention. However we believe that the use of {\it stochastic
Cosserat thermo-mechanics} for slender rods in noisy gravitational
fields offers a reliable research tool for the development of the
scenarios outlined in this paper.

\section {Acknowledgements}

The authors are grateful to BAE SYSTEMS (Warton) and the
Leverhulme Trust for support and are indebted to valuable
conversations with B Schutz, C Cutler, J Hough, H Ward, M Cruise
and C Speake.

\newpage

\begin{figure}
\epsfig{file=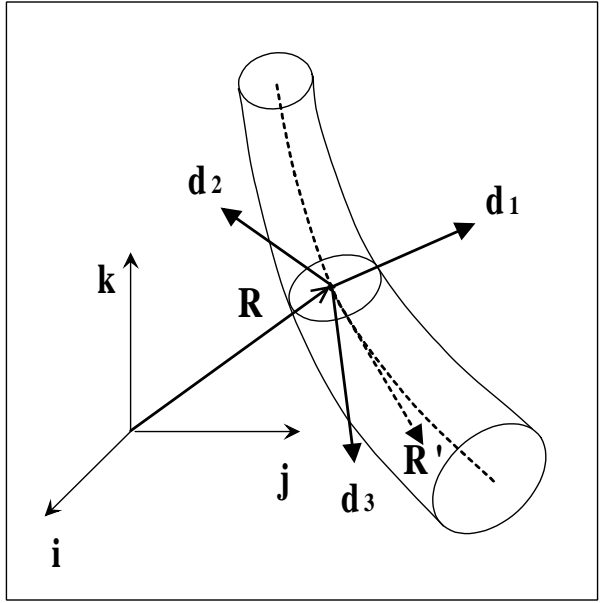,width=10cm,angle=0}
\caption{Segment of a rod and the vectors that enter into its
Cosserat description. } \label{rodfig}
\end{figure}

\begin{figure}
\epsfig{file=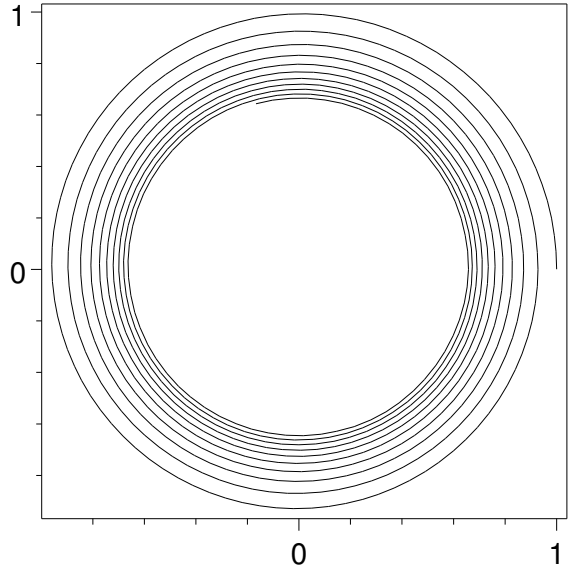,width=9cm,angle=90}
\caption{Image of a spiral with $\mu = 0.3$ and $\ell = 16\,\pi$
plotted in the $\frac{x}{R}$ - $\frac{y}{R}$ plane. The radius of
curvature of this spiral varies from $1/\kappa_0(0) = R$ at the
outer end to $1/\kappa_0(L) = 0.65\,R$ at the inner end with
approximately 8 ($\approx \ell/2\pi$) windings. }
\label{fig_spiral}
\end{figure}

\begin{figure}
\epsfig{file=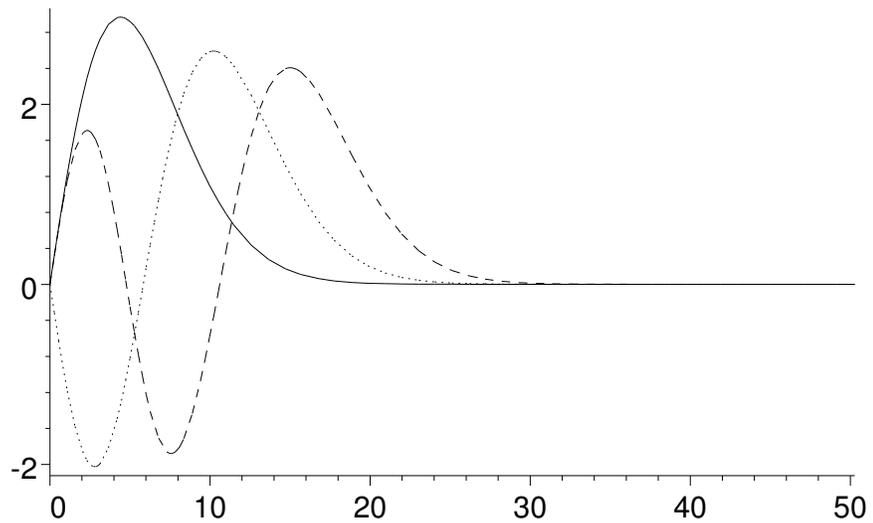,width=8cm,angle=90} \caption{Behaviour of the
first 3 eigen-functions,  $\psi_1(s)$ (solid), $\psi_2(s)$
(dotted) and $\psi_3(s)$ (dashed), for a spiral with $\mu = 0.3$
and $\ell = 16\,\pi$, plotted against $\sigma = s/R \in [0,
\ell]$). } \label{fig_psi}
\end{figure}

\begin{figure}
\epsfig{file=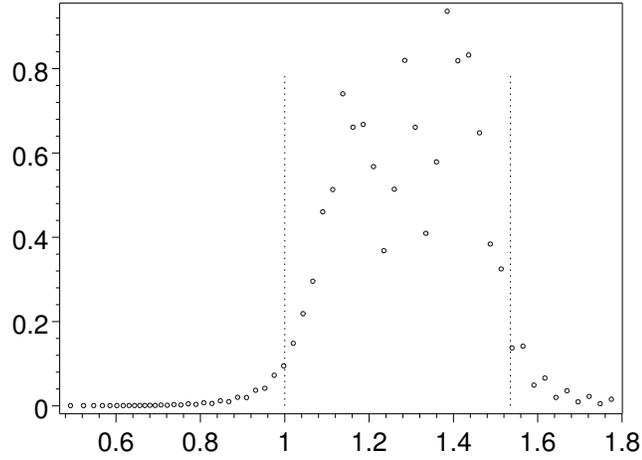,width=8cm,angle=90} \caption{The  overlap
parameters $\hat{a}_p^2 = \ell (a_{1 p}^2 + a_{2 p}^2)$ for the
first 60 normal modes  ($1 \le p \le 60$) of a spiral with $\mu =
0.3$ and $\ell = 16\,\pi$ plotted against the non-dimensional
angular frequency $\hat{\omega}_p = \omega_p/\Omega(0)$ calculated
from $\lambda_p$ using \eqref{omega_p_Omega}. Relatively larger
$\hat{a}_p^2$ are obtained for $30 \lesssim p \lesssim 50$
corresponding to $1 \lesssim \hat{\omega}_p  \lesssim
\kappa_0(\ell)$=1.54 between the two vertical lines.}
\label{fig_aa}
\end{figure}

\end{document}